\documentclass[a4paper]{article}

\usepackage{INTERSPEECH2022}
\usepackage[backref=false]{hyperref}
\usepackage{multirow}
\hypersetup{plainpages = false,
              breaklinks=true,
              linktocpage,
              colorlinks   = true, %Colours links instead of ugly boxes
              urlcolor     = blue, %Colour for external hyperlinks
              linkcolor    = blue, %Colour of internal links
              citecolor   = red, %Colour of citations              pdfauthor={Niko Brummer},
              pdfauthor={Niko Brummer},
              pdfstartpage=1, 
              pdfstartview=FitH,  
              pdfkeywords={}}

\title{Toroidal Probabilistic Spherical Discriminant Analysis}
\name{Anna Silnova$^{1}$,
Niko Br\"ummer$^2$,
Albert Swart$^3$,
Luk\'a\v{s} Burget$^1$
}
\address{
    $^1$Brno University of Technology, Speech@FIT and IT4I Center of Excellence, Brno, Czechia\\
	$^2$Amazon Alexa, South Africa\\
	$^3$Speechly, Finland
	}
\email{isilnova@fit.vutbr.cz, niko.brummer@gmail.com, adswart@gmail.com}

\def\zvec{\mathbf{z}}

\def\vvec{\mathbf{v}}
\def\wvec{\mathbf{w}}
\def\evec{\mathbf{e}}
\def\tvec{\mathbf{t}}

\DeclareMathOperator{\trace}{tr}

\DeclareMathOperator*{\argmax}{argmax}

\def\expv#1#2{\left\langle#1\right\rangle_{#2}}

\def\LR{\text{LR}}
\def\R{\mathbb{R}}

\def\Fmat{\mathbf{F}}

\def\Emat{\mathbf{E}}
\def\Imat{\mathbf{I}}
\def\Ymat{\mathbf{Y}}
\def\Xmat{\mathbf{X}}
\def\Rmat{\mathbf{R}}
\def\Tmat{\mathbf{T}}

\def\Kmat{\mathbf{K}}

\def\Zmat{\mathbf{Z}}

\def\yvec{\mathbf{y}}

\def\nvec{\mathbf{n}}
\def\xvec{\mathbf{x}}

\def\rvec{\mathbf{r}}

\def\muvec{\boldsymbol{\mu}}

\def\avec{\mathbf{a}}
\def\bvec{\mathbf{b}}

\def\nulvec{\boldsymbol{0}}

\def\const{\text{const}}

\def\S{\mathbb{S}}
\def\ellvec{\boldsymbol{\ell}}

\newcommand{\inv}[1]{%  
\ifx{#1}{\Imat}           %this test fails
  #1
\else
  {#1}^{-1}
\fi
}

\def\bmat#1{\begin{bmatrix}#1\end{bmatrix}}

\def\norm#1{\left\|#1\right\|}

\usepackage{tikz}
\usetikzlibrary{bayesnet}

\usetikzlibrary{arrows,shapes,backgrounds,positioning,fit}

\tikzstyle{cbox} = [rectangle,draw=blue!100,thick,align=center,rounded corners = 3pt]
\tikzstyle{lbox} = [rectangle,draw=blue!100,thick,align=left,rounded corners = 3pt]
\tikzstyle{ccircle} = [circle,draw=blue!100,thick,align=center,inner sep = 0]
\tikzstyle{ctext} = [rectangle,align=center,inner sep = 4pt]
\tikzstyle{ltext} = [rectangle,align=left,inner sep = 4pt]
\tikzstyle{solder} = [circle,draw,fill,inner sep = 0, minimum size = 3pt]

\def\textit#1{\emph{#1}}

\def\textit#1{\emph{#1}}

\def\vmf{\mathcal{V}}

\begin{document}

\maketitle
\begin{abstract}
In speaker recognition, where speech segments are mapped to embeddings on the unit hypersphere, two scoring back-ends are commonly used, namely cosine scoring and PLDA. We have recently proposed PSDA, an analog to PLDA that uses Von Mises-Fisher distributions instead of Gaussians. In this paper, we present toroidal PSDA (T-PSDA). It extends PSDA with the ability to model within and between-speaker variabilities in toroidal submanifolds of the hypersphere. Like PLDA and PSDA, the model allows closed-form scoring and closed-form EM updates for training. On VoxCeleb, we find T-PSDA accuracy on par with cosine scoring, while PLDA accuracy is inferior. On NIST SRE'21 we find that T-PSDA gives large accuracy gains compared to both cosine scoring and PLDA.\footnote{The contributions of Niko Br\"ummer and Albert Swart to this paper were performed while they were employed by Phonexia and before they joined Amazon and Speechly, respectively.} 
% We show how the self-conjugacy of this distribution gives closed-form likelihood-ratio scores, making it a drop-in replacement for PLDA at scoring time. All kinds of trials can be scored, including single-enroll and multi-enroll verification, as well as more complex likelihood ratios that could be used in clustering and diarization. Learning is done via an EM-algorithm with closed-form updates. We explain the model and present some first experiments.
\end{abstract}
\noindent\textbf{Index Terms}: speaker recognition, PSDA, Von Mises-Fisher

\section{Introduction}  
Probabilistic \textit{linear} discriminant analysis (PLDA)~\cite{PLDA-IOFFE,PLDA-Prince}, is a popular back-end for scoring speaker recognition embeddings (e.g., i-vectors~\cite{ivec} or x-vectors~\cite{snyder2017}) in $\R^D$, following~\cite{Kenny_HTPLDA,NikoOdyssey10}. However,~\cite{Dani_length_norm} showed that length-normalizing the embeddings onto the unit sphere, $\S^{D-1}$ has a Gaussianizing effect that improves speaker verification performance, and this has been the standard practice ever since. One disadvantage of the length-normalization is that within-speaker variability is squashed in the radial direction, making it \textit{speaker-dependent}, in violation of the PLDA assumption of a constant within-class distribution. Moreover, given a flexible, discriminatively trained embedding extractor, it is often found that cosine scoring (dot products between embeddings on the hypersphere) outperforms PLDA, especially when the test data is in-domain, e.g.,~\cite{SpeakIn_VoxSRC21,zeinali2019but}. 

In our previous work~\cite{brummer22_interspeech}, we introduced \textit{probabilistic spherical discriminant analysis} (PSDA), where the observed and hidden variables have Von Mises-Fisher (VMF) rather than Gaussian distributions. We found the performance of the PSDA model to be very similar to cosine scoring because it has very limited modeling capacity due to the low amount of trainable parameters.  This paper presents an extended version of the PSDA model---\textit{toroidal PSDA} (T-PSDA), where the observed data still live on the unit sphere and have VMF distributions, but now we have added a structured space (defined via a larger set of trainable parameters) where the hidden variables live. An open-source implementation of T-PSDA is available.\footnote{\url{https://github.com/bsxfan/Toroidal-PSDA}} 

\section{Theory}
\subsection{The Von Mises-Fisher distribution}
When embeddings in Euclidean space, $\R^D$ are length-normalized, they are projected onto the \emph{unit hypersphere}:%\footnote{Do not confuse \textit{sphere} with \textit{ball}: $\S^{d-1}$ is the surface of the ball. Euclidean norm is denoted $\norm{\xvec}=\sqrt{\xvec'\xvec}$.}
\begin{align}
\S^{D-1} &= \{\xvec\in\R^D:\,\norm{\xvec}=1\},    
\end{align}
where we model them with the Von-Mises Fisher (VMF) distribution, which for $\xvec\in\S^{D-1}$ has the density~\cite{MardiaJupp}: 
\begin{align}
\label{eq:vmf_1}
\vmf(\xvec\mid\muvec, \kappa) &= \frac{C_\nu(\kappa)}{(2\pi)^{d/2}}e^{\kappa\muvec'\xvec},
\end{align}
where $C_\nu(\kappa) =\frac{\kappa^{\nu}}{I_\nu(\kappa)}$ and $\nu=\frac{D}{2}-1$ and $I_\nu(\kappa)$ is the modified Bessel function of the first kind. The parameters are the \emph{mean direction}, $\muvec\in\S^{D-1}$ and the \emph{concentration}, $\kappa\ge0$ . In terms of the \textit{natural parameter}, $\avec=\kappa\muvec$, the VMF density can alternatively be expressed as:
\begin{align}
\label{eq:vmf_2}
\vmf(\xvec\mid\avec) &= \frac{\bar C_{\nu}(\kappa)}{(2\pi^{d/2})}e^{\avec'\xvec-\kappa},
\end{align}
where $\kappa=\norm{\avec}$ and $\bar C_{\nu}(\kappa) = \frac{e^{\kappa}\kappa^{\nu}}{I_\nu(\kappa)}$.

\subsection{T-PDSA model definition}
T-PSDA is inspired by the original (full) PLDA model that has both a hidden speaker factor and a hidden within-speaker (channel) factor. We replicate this structure with T-PSDA. However, unlike in PLDA, the spherical geometry allows \textit{multiple} speaker factors and \textit{multiple} channel factors. For \textit{each} speaker, let there be a set of $m\ge1$ hidden \emph{speaker factors}, $\Zmat=\{\zvec_i\}_{i=1}^m$, where $\zvec_i\in\S^{d_i-1}$, where all $d_i\ge1$.\footnote{We include the degenerate case, $d_i=1$, where $\S^0=\{-1,1\}$.} We represent $\zvec_i\in\R^{d_i}$, subject to $\zvec_i'\zvec_i=1$. Let $n\ge m$ and with every observation $t$, let there be associated a set of $n-m$ hidden \emph{within-speaker factors}, $\Ymat_t=\{\yvec_{ti}\}_{i=m+1}^n$, where $\yvec_{ti}\in\S^{d_i-1}$ and all $d_i\ge1$. (If $m=n$, it is understood that there are no within-speaker factors.) We define $D_s=\sum_{i=1}^n d_i$ and restrict $D_s\le D$. For an observation, $\xvec_t\in\S^{D-1}$, we define the VMF likelihood:
\begin{align}
\label{eq:defTPSDA}
\begin{split}
    P(\xvec_t\mid\Zmat,\Ymat_t) &= \vmf(\xvec_t\mid\muvec_t,\kappa), \\
\end{split}
\end{align}
where $\kappa>0$ represents the unstructured component of the within-speaker variability. Structured within and between-speaker variabilities are obtained by letting the mean direction, $\muvec_t\in\S^{D-1}$, be a linear combination of the hidden variables:\footnote{Attention: $i$ is \textit{not} a speaker index. All of the variables $\Zmat=\{\zvec_i\}_{i=1}^m$ represent a \textit{single} speaker.}
\begin{align}
\label{eq:mu}
    \muvec_t&=\sum_{i=1}^m w_i\Kmat_i\zvec_i \,+ \sum_{i=m+1}^n w_i\Kmat_i\yvec_{ti}.
\end{align}
By restricting the \textit{weights} as $\sum_{i=1}^{n} w_i^2=1$ and the \textit{factor loading matrices}, $\Kmat_i\in\R^{D\times d_i}$, as $\Kmat_i'\Kmat_i=\Imat_{d_i}$ and $\Kmat'_i\Kmat_{j\ne i}=\nulvec$, we ensure $\muvec_t'\muvec_t=1$. This linear combination spans a $D_s$-dimensional \textit{linear subspace}, but since there are $n$ restrictions of the forms $\zvec'_i\zvec_i=1$ and $\yvec'_{ti}\yvec_{ti}=1$, $\muvec_t$ is in fact restricted to submanifold of dimension $D_s-n$. In the special case that all $d_i=2$, this manifold is known as the \textit{Clifford torus}---in the general case, we use the term \textit{toroidal manifold} and use it to name the model \textit{toroidal PSDA} (T-PSDA).

For given observations, their dimensionality $D$ is fixed, but we have some freedom to choose the hidden variable structure. We can choose the linear subspace dimension, $D_s$, and the number of factors, $n$, and their dimensions, $d_i$. We would choose $D_s<D$ if we believe the data lives close to a \emph{linear} subspace of $\R^D$. Moreover, the more factors ($n$ of them) we choose, the `thinner' the toroidal manifold becomes because it is of dimension $D_s-n$. However, the VMF likelihood~\eqref{eq:defTPSDA} allows the data to stray away from $\muvec_t$ and appear anywhere on $\S^{D-1}$. The VMF concentration, $\kappa$, controls how far the data can stray from the toroid where $\muvec_t$ lives. 

If we set $n=m=1$ and $d_1=D$, then T-PSDA degenerates to the PSDA model of~\cite{brummer22_interspeech}. If additionally, we fix $\gamma_1=0$, the scoring with the resulting model is equivalent to cosine scoring.

\subsection{Hidden variable prior and posterior}
As in PLDA, speakers are independent. For a given speaker with observed data, $\Xmat=\{\xvec_t\}_{t=1}^T$, the associated hidden variables are $\Zmat$ and $\Ymat=\{\Ymat_t\}_{t=1}^T$. The T-PSDA model is completed by specifying a \textit{conjugate prior}:
\begin{align}
P(\Zmat,\Ymat) &= \prod_{i=1}^m \vmf(\zvec_i\mid\vvec_i,\gamma_i) \prod_{t=1}^T\prod_{i=m+1}^n \vmf(\yvec_{ti}\mid\vvec_i,\gamma_i),
\end{align}
where $\vvec_i\in\S^{d_i-1}$ and $\gamma_i\ge0$ are trainable parameters. The posterior is proportional to (and can be recovered from) the joint distribution:
\begin{align*}
\begin{split}
\log P(\Xmat,\Zmat,\Ymat) &= \sum_{i=1}^m \zvec_i'\bigl[\gamma_i\vvec_i + \kappa w_i\Kmat'_i\sum_{t=1}^N\xvec_t \bigr] \\
&+ \sum_{i=m+1}^n \sum_{t=1}^N \yvec_{ti}'\bigl[\gamma_i\vvec_i + \kappa w_i\Kmat'_i\xvec_t\bigr] + \const.
\end{split}
\end{align*}
This shows the posterior remains factorial---\textit{all the hidden variables remain independent}.\footnote{Why do we not get explaining away? After all, $n$ hidden variables are jointly responsible for each $\xvec_t$. This can be understood by the mutually orthogonal factor loading matrices, the $\Kmat_i$. The hidden factor $i$ is actually solely responsible for the projected observation $\Kmat_i'\xvec_t$.} The posterior factors are:
\begin{align}
\label{eq:zpost}
&P(\zvec_i\mid\Xmat) = \vmf(\zvec_i\mid\tilde\vvec_i), \;\;\;\;
\tilde\vvec_i=\gamma_i\vvec_i + \kappa w_i\Kmat'_i\sum_{t=1}^T\xvec_t, \\
\label{eq:ypost}
&P(\yvec_{ti}\mid\xvec_t) = \vmf(\yvec_{ti}\mid\tilde\vvec_{ti}), \;\;\;\;
\tilde\vvec_{ti}=\gamma_i\vvec_i + \kappa w_i\Kmat'_i\xvec_t,
\end{align}
where we have used the natural VMF parametrization of~\eqref{eq:vmf_2}.

\subsection{Scoring}
Because of the conjugacy, the hidden variables can be integrated out using the candidate's trick~\cite{Besag1989ACF}:
\begin{align}
\label{eq:marglh1}
P(\Xmat) 
= \frac{P(\Xmat\mid\Zmat_0,\Ymat_0)P(\Zmat_0)P(\Ymat_0)}
{P(\Zmat_0\mid\Xmat)P(\Ymat_0\mid\Xmat)},
\end{align}
where $\Zmat_0$ and $\Ymat_0$ are any convenient values for the hidden variables. Using this for two sets of observations, $\Emat$ and $\Tmat$, the likelihood ratio (LR) between the \textit{same-speaker} and \textit{different-speakers} hypotheses can be expressed as:
\begin{align}
\LR &= \frac{P(\Emat,\Tmat)}{P(\Emat)P(\Tmat)} 
= \frac{P(\Zmat_0\mid\Emat)P(\Zmat_0\mid\Tmat)}{P(\Zmat_0\mid\Emat,\Tmat)P(\Zmat_0)},
\end{align}
where factors involving $\Ymat_0$ cancel. By denoting $\Zmat_0=\{\zvec_i\}_{i=1}^m$ and plugging in~\eqref{eq:zpost}, the LR becomes:
\begin{align*}
\begin{split}
\LR&= \prod_{i=1}^m \frac{\vmf(\zvec_i\mid\gamma_i\vvec_i + \kappa w_i\Kmat'_i\tilde\evec)\,
\vmf(\zvec_i\mid\gamma_i\vvec_i + \kappa w_i\Kmat'_i\tilde\tvec)}{\vmf(\zvec_i\mid\gamma_i\vvec_i + \kappa w_i\Kmat'_i(\tilde\evec+\tilde\tvec))\,\vmf(\zvec_i\mid\gamma_i\vvec_i)} \\
&= \prod_{i=1}^m \frac{\vmf(\zvec_i\mid\ellvec_i)\,\vmf(\zvec_i\mid\rvec_i)}{\vmf(\zvec_i\mid\bvec_i)\,
\vmf(\zvec_i\mid\nvec_i)} .
\end{split}
\end{align*}
Here, $\tilde\evec$ and $\tilde\tvec$ are the sums of the observations in respectively $\Emat$ and $\Tmat$. We have introduced mnemonic short-hand notation for the parameters of the above VMF factors: $\ellvec$ for left, $\rvec$ for right, $\bvec$ for both, and $\nvec$ for none.  In the score, all factors of the form $e^{\avec'\zvec_i}(2\pi)^{-d_i/2}$ cancel---see~\eqref{eq:vmf_2}, so that the score simplifies to: 
\begin{align}
\LR &= \prod_{i=1}^m\frac{\bar C_{\nu_i}(\norm{\ellvec_i})e^{-\norm{\ellvec_i}}\, \bar C_{\nu_i}(\norm{\rvec_i})e^{-\norm{\rvec_i}}}
{\bar C_{\nu_i}(\norm{\bvec_i})e^{-\norm{\bvec_i}}\,\bar C_{\nu_i}(\norm{\nvec_i})e^{-\norm{\nvec_i}}}.
\end{align}
To get an idea of what the score does, let us assume the ideal situation where the $\gamma_i=0$, so that the speaker factors have uniform distributions. Noting that $\bar C_\nu(x)$ is almost constant compared $e^{-x}$ on the scale that $x$ typically varies, the log LR can be \textit{approximated} as:
\begin{align*}
\label{eq:score_shortcut}
\kappa\sum_{i=1}^m \lvert w_i\rvert\Bigl(\norm{\Kmat_i'(\tilde\evec+\tilde\tvec)} - \norm{\Kmat_i'\tilde\evec} - \norm{\Kmat_i'\tilde\tvec}\Bigr) + \const.
\end{align*} 
The score becomes more positive if $\tilde\evec$ and $\tilde\tvec$ are aligned and more negative if they are not. If $\tilde\evec$, or $\tilde\tvec$, or both are summed over multiple aligned inputs, then the score magnitude can increase. The score is essentially a linear fusion, weighted by the $\lvert w_i\rvert$. The score magnitude is also scaled by the observation VMF concentration, $\kappa$.  

Finally, note that when scoring, we never use the $w_i$ and the $\Kmat_i$ for $i>m$, i.e., for the within-speaker subspaces. It may be asked what is the use of this part  of the model? Yes, that part does not participate in scoring, but it \textit{does} play a part when learning the model parameters. The three sources of variability---the unstructured VMF noise (parametrized by $\kappa$) and the structured within and between-speaker variabilities--- \textit{compete} to explain the total variability in the observed data.    

\subsection{Training}
For a given T-PSDA architecture, the parameters are $\kappa$ and $\{\Kmat_i,w_i,\vvec_i,\gamma_i\}_{i=1}^n$ and they can be learned with maximum likelihood, using an EM algorithm. The E-step is to compute the \textit{EM auxiliary}, i.e.\ the log-likelihood expectation w.r.t.\ the hidden variable posterior:
\begin{align*}
    Q = \sum_{s=1}^S \expv{\log P(\Xmat_s\mid\Zmat,\Ymat) + \log P(\Zmat,\Ymat)}{P(\Zmat,\Ymat\mid\Xmat_s)},
\end{align*}
where the data for speaker $s$ is $\Xmat_s=\{\xvec_{st}\}_{t=1}^{T_s}$, and $S$ is the total number of speakers. Recall that the posterior is a product of VMF factors given by~\eqref{eq:zpost} and~\eqref{eq:ypost}. Since the log-likelihood is linear in the hidden variables, we can simply plug in the \textit{posterior expectations}, $\bar\zvec_{si}$ and $\bar\yvec_{sti}$, in place of the hidden variables:
\begin{align}
\begin{split}
Q&=  Q^{(x)} + \sum_{i=1}^m Q^{(z)}_i + \sum_{i=m+1}^n Q^{(y)}_i,
\end{split}
\end{align}
where
\begin{align}
\begin{split}
\label{eq:q}
Q^{(x)}&=\sum_{s=1}^S\sum_{t=1}^{T_s} \log\vmf(\xvec_{st}\mid\bar\muvec_{st},\kappa), \\
Q^{(z)}_i&=\sum_{s=1}^S\log \vmf(\bar\zvec_{si}\mid\vvec_i,\gamma_i), \\
Q^{(y)}_i&=\sum_{s=1}^S \sum_{t=1}^{T_s} \log\vmf(\bar\yvec_{sti}\mid\vvec_i,\gamma_i), \\
\end{split}
\end{align}
where we have defined:
\begin{align}
    \bar\muvec_{st} &= \sum_{i=1}^m w_i\Kmat_i\bar\zvec_{si} \,+ \sum_{i=m+1}^n w_i\Kmat_i\bar\yvec_{sti}.
\end{align}
The M-step maximizes $Q$ w.r.t.\ all the parameters. We can maximize the above $Q$-terms independently. For the $Q_i^{(z)}$ and $Q_i^{(y)}$, these are standard VMF maximum likelihood problems:
\begin{align}
\begin{split}
    \vvec_i,\gamma_i &\gets \argmax_{\vvec,\gamma} \prod_{s=1}^S \vmf(\bar\zvec_{si}\mid\vvec,\gamma),\;\;\; i \le m,\\
    \vvec_i,\gamma_i &\gets \argmax_{\vvec,\gamma} \prod_{s=1}^S\prod_{t=1}^{T_s} \vmf(\bar\yvec_{sti}\mid\vvec,\gamma),\;\;\; m< i \le n.
\end{split}
\end{align}
It remains to maximize $Q^{(x)}$. First, we can maximize it w.r.t.\ $\{\Kmat_i,w_i\}$, independently of $\kappa$, which is equivalent to maximizing:
\begin{align*}
\begin{split}
%&\argmax_{\{\Kmat_i,w_i\}}Q^{(x)}=\\
%&\argmax_{\{\Kmat_i,w_i\}}
&\sum_{s=1}^S \sum_{t=1}^{T_s}\biggl[ \sum_{i=1}^m w_i\xvec'_{st}\Kmat_i\bar\zvec_{si} \,+ \sum_{i=m+1}^n w_i\xvec'_{st}\Kmat_i\bar\yvec_{sti}\biggr]\\
& =\sum_{i=1}^m w_i \trace\bigl(\Kmat_i\sum_{s}\bar\zvec_{si}\sum_t\xvec'_{st}\bigr)\,+ 
\sum_{i=m+1}^n w_i \trace\bigl(\Kmat_i\sum_{st}\bar\yvec_{sti}\xvec'_{st}\bigr) \\
&=\sum_{i=1}^n w_i\trace(\Kmat_i\Rmat'_i),
\end{split}
\end{align*}
where $\Rmat_i$ is defined by matching the final line to the one above. We do a few iterations of co-ordinate ascent, alternatively updating $\wvec=(w_1,\ldots,w_n)$ and $\Fmat=\bmat{\Kmat_1&\cdots&\Kmat_n}$. When $\Fmat$ is temporarily fixed, we define $\tilde\wvec=(\tilde w_1,\ldots,\tilde w_n)$, where $\tilde w_i=\trace(\Kmat_i\Rmat'_i)$. Conversely, when $\wvec$ is temporarily fixed, we define $\tilde\Fmat=\bmat{w_1\Rmat_1&\cdots&w_n\Rmat_n}$. The maximizing updates, subject to the constraints $\wvec'\wvec=1$ and $\Fmat'\Fmat=\Imat$, are:\footnote{The $\Fmat$ update requires the symmetric, positive-definite matrix square root.}
\begin{align}
\wvec&\gets\frac{\tilde\wvec}{\norm{\tilde\wvec}} &&\text{and}&
\Fmat&\gets\tilde\Fmat(\tilde\Fmat'\tilde\Fmat)^{-\frac12}.
\end{align}
Finally, once the optimal values for $\wvec$ and $\Fmat$ have been obtained, we can fix them to find the optimal $\kappa$:
\begin{align}
\begin{split}
\kappa\gets &\argmax_{\kappa} \sum_{st}\log\vmf(\xvec_{st}\mid\bar\muvec_{st}) \\
=&\argmax_{\kappa} T\log\frac{\kappa^\nu}{I_{\nu}(\kappa)} + \kappa\sum_{i=1}^n w_i\trace(\Kmat_i\Rmat'_i),
\end{split}
\end{align}
where $T=\sum_s T_s$ and $\nu=\frac{D}2-1$. This scalar optimization can be done with a general-purpose (typically derivative-free) numerical optimization algorithm.% Finally, to check and monitor the EM algorithm, it is useful to print out the ML objective, the \emph{marginal log-likelihood}: 

\section{Experiments}
We perform the experiments with the T-PSDA model on two datasets: NIST SRE'21~\cite{nist_sre21} and VoxCeleb~\cite{Nagrani19,chung2018voxceleb2}. In both cases, we compare the speaker verification performance of the proposed model with cosine scoring and PLDA. Experiments with the original PSDA are not repeated here, because it performs on par with cosine scoring~\cite{brummer22_interspeech}.

\subsection{NIST SRE'21}
 For the experiments on the NIST SRE'21 evaluation set, we use exactly the same ResNet152 embedding extractor as used in~\cite{BUTOdyssey22}. We used the version of the extractor that was fine-tuned on long (10\,s) segments. This embedding extractor was used as input for a number of different scoring back-ends that we compare below. 

The PLDA and T-PSDA models require training, while cosine scoring does not. All back-end models were trained on the full NIST CTS superset~\cite{sre_cts_superset} unlike in~\cite{BUTOdyssey22} where only the English, Mandarin, and Cantonese subsets were used (hence, there are minor performance differences between the results of this paper and those presented in~\cite{BUTOdyssey22}). In all cases, the embeddings were centered and length-normalized. We optionally used linear discriminant analysis (LDA) to reduce the dimensionality of the embeddings from 256 to 100, because we have observed before~\cite{BUTOdyssey22} that roughly half of the embedding dimensions have almost no useful variability and that LDA improved some of the back-ends. The parameters for LDA and centering were obtained from the same training set.
\begin{table}[!htb]
  \caption{
 Performance of the baseline  and T-PSDA back-ends with and without score normalization on evaluation set of NIST SRE'21. The performance metrics are minimum cost and equal-error-rate (EER,\,\%) as computed by the NIST scoring tool.} 
  \label{tab:results:baseline}
  \centering
 \begin{tabular}{ c l  c c  c c }
    \toprule
    %& \multicolumn{6}{c}{\textbf{minDCF}} \\
    && \multicolumn{2}{c}{\textbf{no S-norm}}&\multicolumn{2}{c}{\textbf{S-norm}}\\
     & \textbf{} & 
     %\multicolumn{2}{c}{std embd}   \\
    %\cmidrule(lr){3-4} \cmidrule(lr){5-6} \cmidrule(lr){7-8}
      \textbf{min\_C}  & \textbf{EER}& \textbf{min\_C}  & \textbf{EER}  \\
    \midrule
1 &cos & 0.490&9.18 & 0.520&8.62  \\ 
2 &cos + LDA& 0.468& 8.10	&0.596 &7.77  \\ 
3 & PLDA &0.444&7.88&0.653&7.55  \\ 
4 & PLDA + LDA&0.452 &7.78&0.650&7.80  \\ 
5& T-PSDA&\textbf{0.381}& \textbf{6.16}&\textbf{0.375} &\textbf{5.77} \\
%5 & PLDA (diag W) &0.478&8.68&0.487&8.2  \\ 		
%6 & PLDA + LDA (diag W)&0.461 &7.97&0.636&7.81  \\ 			
    \bottomrule
  \end{tabular}
%  \vspace{-4mm}
\end{table}
\begin{table*}[!htb]
  \caption{Performance of the baseline and T-PSDA back-ends. The performance is reported in terms of equal-error-rate (EER,\,\%) and minimum Detection Cost Function  computed and target probability $p_{\mathrm{tar}}=0.05$ ($\mathrm{min\_C}$).} 
  \label{tab:results_vox:baseline}
  \centering
%  \begin{tabular}{ c l  c c  c c c c }
%     \toprule
%     %& \multicolumn{6}{c}{\textbf{minDCF}} \\
%     && \multicolumn{2}{c}{\textbf{Vox1-O}}&\multicolumn{2}{c}{\textbf{Vox1-H}}&\multicolumn{2}{c}{\textbf{SITW}}\\
%      & \textbf{} & 
%      %\multicolumn{2}{c}{std embd}   \\
%     %\cmidrule(lr){3-4} \cmidrule(lr){5-6} \cmidrule(lr){7-8}
%       \textbf{min\_C}  & \textbf{EER}& \textbf{min\_C}  & \textbf{EER} & \textbf{min\_C}  & \textbf{EER} \\
%     \midrule
% 1 &cos &0.072& 1.09&	0.126&	2.15 &	0.073&1.32	 \\ 
% 2 &cos + S-norm &\textbf{0.061}& 1.01&	\textbf{0.118}&\textbf{2.03} &	\textbf{0.070}&1.28	 \\ 
% 3 &cos +LDA &0.116& 1.54&	0.269&	4.34 &	0.103&1.63	 \\ 	
% 4 &PLDA &0.254&4.47&0.325&6.80&0.202& 3.94 \\
% 5 &PLDA +LDA&0.113& 1.55&0.198	&3.44&0.088	&1.48	 \\ 
% 6& T-PSDA  &0.069& 1.07&	0.127&	2.16 &	0.075&	1.32 \\ 		
% 7& T-PSDA + S-norm  &0.065& \textbf{0.97}&	0.119&	2.05 &	0.071&	\textbf{1.27} \\ 

 \begin{tabular}{ c l  c c  c c c c c c }
    \toprule
    %& \multicolumn{6}{c}{\textbf{minDCF}} \\
    && \multicolumn{4}{c}{\textbf{no S-norm}}&\multicolumn{4}{c}{\textbf{S-norm}}\\
    && \multicolumn{2}{c}{\textbf{Vox1-O}}&\multicolumn{2}{c}{\textbf{Vox1-H}}&\multicolumn{2}{c}{\textbf{Vox1-O}}&\multicolumn{2}{c}{\textbf{Vox1-H}}\\
     & \textbf{} & 
     %\multicolumn{2}{c}{std embd}   \\
    %\cmidrule(lr){3-4} \cmidrule(lr){5-6} \cmidrule(lr){7-8}
      \textbf{min\_C}  & \textbf{EER}& \textbf{min\_C}  & \textbf{EER} & \textbf{min\_C}  & \textbf{EER}& \textbf{min\_C}  & \textbf{EER} \\
    \midrule
1 &cos &0.071& 1.10&	\textbf{0.126}&	\textbf{2.13} &	\textbf{0.061}& 1.01&	\textbf{0.118}&\textbf{2.03}	 \\ 

2 &cos +LDA &0.116& 1.54&	0.263&	4.15 &	0.100&1.38& 0.160&2.62 \\ 	
3&PLDA &0.254&4.47&0.325&6.80&0.213	&3.48&0.285 &6.05\\
4 &PLDA +LDA&0.113& 1.55&0.198	&3.44&	0.106&1.40& 0.152&	2.96 \\ 
5& T-PSDA  &\textbf{0.069}&\textbf{1.07}&	0.127&	2.16 &	 0.065& \textbf{0.97}&	0.119&	2.05 \\ 		
% 7& T-PSDA + S-norm  &0.065& \textbf{0.97}&	0.119&	2.05 &	0.071&	\textbf{1.27} \\ 

%2 &cos + LDA& 0.42& 8.65	&0.40 & 7.53 \\ 
%3 & PLDA &0.51&11.88&&  \\ 
%4 & PLDA + LDA &0.39&8.41&&  \\ 
    \bottomrule
  \end{tabular}
%  \vspace{-4mm}
\end{table*}

Table~\ref{tab:results:baseline} compares the proposed T-PSDA against the two baseline models (cosine scoring and PLDA), with and without LDA and with and without adaptive score normalization~\cite{matejka2017analysis}. For PLDA we set the sizes of each of the speaker and channel subspaces to 100.\footnote{For PLDA after LDA down to 100 dimensions, the subspaces are therefore of full rank, so that PLDA degenerates to the \textit{two-covariance} variant~\cite{NikoOdyssey10}.} For score normalization,  we used the 400 highest scores of the enrollment and test segments against 5000 randomly chosen embeddings from the training set.  
The results of the baseline models are shown in lines 1 to 4 of Table~\ref{tab:results:baseline}.  Notice that score normalization leads to significant performance degradation in terms of minimum cost for all baseline back-ends, while in some cases, it improves equal error rate (EER). Also, let us notice that in this experiment (NIST'21), using LDA is beneficial for cosine scoring, but not for PLDA. For the other experiment (VoxCeleb, described below), we observe the \textit{opposite}.  

T-PSDA provides great freedom in selecting its parameters: the number and sizes of the hidden speaker and channel variables. For this reason, it is not feasible to consider all possible combinations of T-PSDA settings to select the best one. We used uniform hidden variable priors: we set all $\gamma_i=0$ and did not learn the prior parameters. For the other parameters, we used the following strategy: we start from the simplest configuration with a single speaker variable and no channel variables. Gradually, one parameter at a time, we increase the complexity of the model. First, we optimize the dimensionality of the single speaker variable%~(see Table~\ref{tab:results:dz} for these results)
; then we experiment with having several speaker variables such that their summed dimensions are the same as what we found optimal in the first step%~(see Table~\ref{tab:results:m}) 
. After this, having the speaker variable dimensions fixed, we introduce channel variability and similarly optimize the number and size of the channel variables%~(Tables~\ref{tab:results:dy} and~\ref{tab:results:n})
. Following this approach, we arrive at the optimal architecture of T-PSDA model: we fix the number of hidden speaker variables to one ($m=1$) and the dimensionality of this variable ($d_1=60$); also we use two 5-dimensional hidden channel variables ($n=3,\;d_2=d_3=5$). The performance of the final model is shown in line 5 of Table~\ref{tab:results:baseline}.

Comparing T-PSDA to the baselines, we observe significant performance improvement in both cases: when score normalization is performed or not. Also, notice that T-PSDA benefits from score normalization, unlike the baselines. However, it is important to mention that these results are achieved with a model found by a greedy search in the space of the T-PSDA configurations. T-PSDA  performance greatly depends on the correct settings for the number and dimensionality of the hidden variables; in some experiments with different configurations of the model, the performance was considerably worse. 
\subsection{VoxCeleb}

For the Audio from Video (AfV) data experiments, we replicate the experimental setup of~\cite{brummer22_interspeech} where a ResNet34 embedding extractor trained on the development part of the VoxCeleb2 dataset was used, and the performance was tested on the original test set of VoxCeleb1 (Vox1-O) and a set of ``hard'' trials constructed out of the whole VoxCeleb1 (Vox1-H). %and core-core condition of Speakers in the Wild (SITW). 
The performance is evaluated in terms of EER and minimum Detection Cost Function with the probability of the target trial set to 0.05.

The results of the baseline models are shown on lines 1 to 4 of Table~\ref{tab:results_vox:baseline}. The results for cosine scoring and the PLDA model are shown with and without LDA reducing the dimensionality of the embeddings from 256 to 200. When the PLDA model is trained on the raw embeddings without dimensionality reduction, we set the sizes of each of the speaker and channel subspaces to 100. When LDA is applied, we train the PLDA with full-rank within and across-class covariances. Notice that LDA  is critical for a good performance of the PLDA model, while for cosine distance scoring, it is rather detrimental. %Also, for the best-performing baseline -- cosine distance scoring, we show the performance with adaptive score normalization (line 2 of the table).
Also, for all baselines, we show the performance with adaptive score normalization. Score normalization is done the same way as for SRE experiments: we use the highest 400 scores of the trial sides against 5000 embeddings from the training set. 

To find the appropriate architecture of the T-PSDA model, we followed an approach similar to the one used in NIST SRE'21 experiments: looking for one configuration parameter at a time and treating the others as fixed. In this way, we found the optimal architecture of the T-PSDA model which is having a single 120-dimensional speaker variable, i.e., $m=1, \;d_1=120$, and five 1-dimensional hidden channel variables ($n=6,\;d_2=\ldots=d_6=1$).
The performance of this model is shown on line 5 of Table~\ref{tab:results_vox:baseline}. For this system, we provide the results with and without score normalization. As seen from the results, for AfV data, T-PSDA performs on par with the best-performing baseline method (cosine scoring), outperforming PLDA.% 
\section{Conclusion}
% We have introduced a model using VMF distributions to model length-normalized speaker embeddings. The model utilizes two sets of hidden variables living on a toroidal submanifold of the observed space. As we have shown, the T-PSDA model allows probabilistic interpretation for the produced scores, hence, it allows for, e.g., domain adaptation. Besides, fast training and scoring recipes are available for T-PSDA. This makes T-PSDA an attractive alternative to a widely used cosine distance scoring and PLDA model. As the experiments indicate, in some cases, using T-PSDA provides a significant performance improvement, while in others, it performs similarly to the alternative scoring strategies.
We have generalized the simple PSDA model of~\cite{brummer22_interspeech} to a new model called T-PSDA. The PSDA model is too simple: lacking trainable parameters, it was not able to outperform cosine scoring in a situation where a trainable PLDA model was able to outperform cosine scoring in a new domain (NIST SRE'21). The T-PSDA model has a set of trainable parameters of a size similar to PLDA. Like PLDA, T-PSDA can model within and between-speaker variabilities in subspaces, while T-PSDA has the advantage of using VMF distributions that better model length-normalized embeddings. These benefits gave a clear performance advantage for T-PSDA on the NIST data. In the VoxCeleb experiment, where the embedding extractor is trained on in-domain data, T-PSDA does not benefit from its domain adaptation capability and performs on par with cosine scoring. In contrast, PLDA performs worse on VoxCeleb.

\section{Acknowledgements}

The work was supported by Czech National Science Foundation (GACR) project NEUREM3 No. 19-26934X, and Horizon 2020 Marie Sklodowska-Curie grant ESPERANTO, No. 101007666. Computing on IT4I supercomputer was supported the Ministry of Education, Youth and Sports of the Czech Republic through the e-INFRA CZ (ID:90140).

\bibliographystyle{IEEEtran}
\bibliography{mybib}

\end{document}